\pdfoutput=1

\documentclass[11pt]{article}

\usepackage[preprint]{acl}

\usepackage{times}
\usepackage{latexsym}
\usepackage{booktabs}
\usepackage[T1]{fontenc}

\usepackage[utf8]{inputenc}

\usepackage{microtype}

\usepackage{inconsolata}
\usepackage{comment}
\usepackage{graphicx}

\usepackage{multirow}
\usepackage{fontawesome}
\usepackage{footmisc}

\title{Is He Extroverted? \\Identifying Missing Relevant Personas for Faithful User Simulation}

\author{\textbf{Weiwen Su\textsuperscript{1,3}},
        \textbf{Yuhan Zhou\thanks{Equal contribution.}\textsuperscript{1}},
        \textbf{Zihan Wang\footnotemark[\value{footnote}]\textsuperscript{1}},
        \textbf{Naoki Yoshinaga\textsuperscript{2,3}},
        \textbf{Masashi Toyoda\textsuperscript{2,3}}
        \\
        \textsuperscript{1}The University of Tokyo,
        \textsuperscript{2}Institute of Industrial Science, The University of Tokyo
        \\
        \textsuperscript{3}Institute for Digital Observatory, The University of Tokyo
        \\
            \texttt{\{su-w, yzhou, zwang, ynaga, toyoda\}@tkl.iis.u-tokyo.ac.jp} 
}

\begin{document}
\maketitle
\begin{abstract}


Existing user simulation approaches focus on generating user-like responses in dialogue. They often assume that the provided persona is sufficient for producing such responses, without verifying whether critical personas are supplied. This raises concerns about the validity of simulation results.
To address this issue, we study the task of identifying persona dimensions (\textit{e.g.}, ``\textit{whether the user is price-sensitive}``) that are relevant but missing in simulating a user's reply for a given dialogue context.
We introduce PICQ-drama (constructed from TVShowGuess), a benchmark of context-aware choice questions, annotated with missing persona dimensions whose absence leads to ambiguous user choices. We further design diverse evaluation criteria for missing persona identification.
Benchmarking leading LLMs on our PICQ-drama dataset demonstrates the feasibility of this task. Evaluation across diverse criteria, along with further analyses, reveals cognitive differences between LLMs and humans and highlights the distinct roles of different persona categories in shaping responses.
%
The dataset is available at:

{\centering
\faGithub\,\url{https://github.com/NioHww/PICQ/}\par}

\end{abstract}

\section{Introduction}

User simulation aims to model the behavior of a target user in a hypothetical situation and is commonly studied to predict the user's responses given a dialogue context and additional data to characterize the user.
Recent large language models (LLMs) have greatly expanded its potential, supporting applications such as non-player characters in games~\cite{Park2023GenerativeAgents}, character-based response generation~\cite{shao2023character,wang-etal-2024-rolellm,tu-etal-2024-charactereval}, and opinion dissemination~\cite{gao2023s3socialnetworksimulationlarge}. These simulations often rely on rich personas or interaction history, either manually prepared or derived from external sources (\textit{e.g.}, Wikipedia), without considering practical simulation situations.
\begin{figure}[t]
\centering
\includegraphics[width=0.97\linewidth]{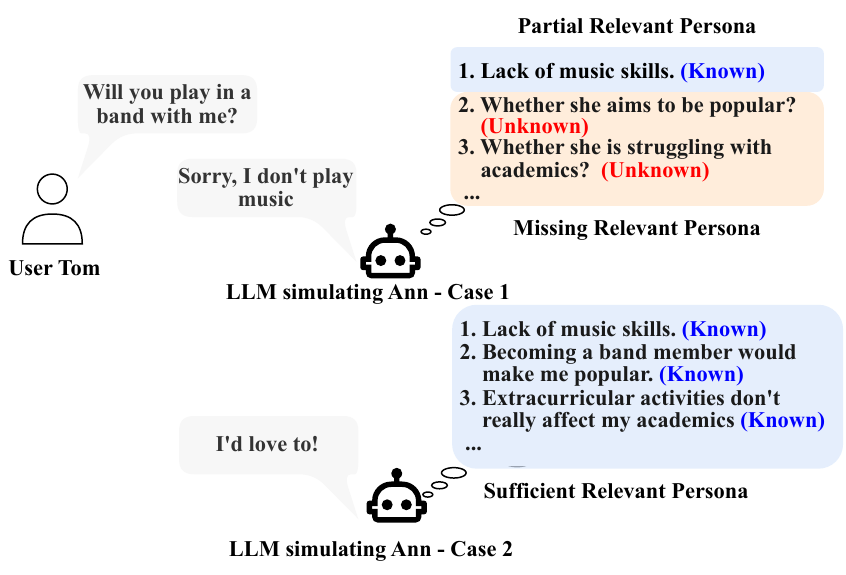}   
\caption{User simulation with sufficient versus partial relevant personas can lead to opposite answers, even when the simulation itself is accurate.}
\label{fig:task}
\end{figure}

Existing user simulation studies attempt to accumulate as comprehensive personas as possible in advance, using external sources such as biographies~\cite{shao2023character,Wang2025OpenCharacterTC} or interviews~\cite{Ng2024HowWC,park2024generativeagentsimulations1000}, to support simulation across a wide range of situations. However, simulations in individual situations are influenced by diverse, situation-specific personas that cannot be fully captured by generic biographies or situation-agnostic interviews. As a result, user simulation based on situation-agnostic comprehensive personas suffers from missing relevant personas (Figure~\ref{fig:task}), leading to unfaithful user simulation. Moreover, since not all personas are relevant to a given situation, using comprehensive personas for a specific situation can be unnecessarily costly.


In this study, to enable faithful user simulation based on relevant personas, we assume a basic persona (including age, gender, and the interlocutor relationship) and focus on identifying missing personas that are relevant in a specific simulation context. By solving this task using LLMs, we ask the following research questions:
\begin{description}
    \item[RQ1:] How well can leading LLMs identify missing relevant personas from context?
    \item[RQ2:] What cognitive patterns emerge in their performance, especially compared to humans?
    \item[RQ3:] Can effective instruction strategies enhance model performance on this task?
\end{description}

To answer these research questions in a controlled and meaningful setting, we construct a new benchmark dataset, PICQ-drama, based on character-rich drama scripts in the TVShowGuess dataset~\cite{sang2022tvshowguess}. We focus on user responses to persona-influenced choice questions (PICQs), where respondents select from a constrained set of options (\textit{e.g.}, ``Would you marry me?''); answers to PICQs naturally reveal user preferences shaped by personas.
Empirically, PICQs constitute the majority of persona-influenced questions (\S~\ref{sec:task}) in the TVShowGuess dataset. 
Our dataset pairs context-aware PICQs with annotated missing relevant persona dimensions. Annotation proceeds in two stages: LLM pre-screening combined with manual verification identifies PICQs from dialogues; annotators determine and describe the missing persona dimensions that influence each PICQ choice. In addition, we propose a multi-faceted evaluation scheme with three metrics, assessing influence on user choices, the difficulty of acquisition (inaccessibility), and alignment to human-annotations. We also design a multi-task instruction strategy designed for this task.


In our experiments, 
we benchmark leading LLMs such as GPT-4.1~\cite{openai2024gpt4technicalreport}, Qwen-3~\cite{yang2025qwen3technicalreport}, and Llama-3.1~\cite{grattafiori2024llama3herdmodels}. The results confirm the feasibility of applying LLMs to this task and validate the effectiveness of our instruction strategy. We evaluate the models from various aspects of influence, inaccessibility, and fidelity with human annotations. We investigate the influence of model scales on the performance, the influence of our instruction strategies, and the cognitive patterns of the models.

Our contributions lie in 
(1) We formulate a new, query-focused task for identifying missing relevant personas to ensure persona sufficiency;
(2) We create the PICQ-drama benchmark with PICQs annotated for missing personas and evaluation metrics;
(3) We design a multi-task instruction strategy to enhance the influence of identified personas;
and (4) We conduct evaluation and analysis to reveal cognitive differences among LLMs and humans.

\section{Related Work}

In this section, we first review the literature on user simulation to position our task.
We then introduce existing persona-augmented dialogue datasets and clarify how our dataset relates to and differs overall.

\subsection{User Simulation}

Recent studies using LLMs to simulate human responses can be categorized by persona granularity: demographic, biography, and individualized personas~\cite{Chen2024FromPT}. 

\smallskip\noindent\textbf{User simulation with demographic persona} captures behavior patterns of certain groups (\textit{e.g.}, ``a 25-year-old white woman'') rather than specific individuals~\cite{deshpande-etal-2023-toxicity,kong-etal-2024-better}. While this setting does not aim to simulate a unique individual, it does not eliminate the issue of persona insufficiency. Specifying only demographic attributes is often insufficient to determine a unique response, and the appropriate output in such cases may be a distribution of plausible behaviors rather than a single prediction.
Therefore, we focus on individual simulations, where the goal is to approximate a specific person's response, making persona sufficiency a critical concern.

\smallskip\noindent\textbf{Individual simulation with biography} mimics fictional characters or celebrities. This type of target often provides abundant persona data for simulation, such as scripts~\cite{tu-etal-2024-charactereval,chen-etal-2023-large}, summarized personas (\textit{e.g.}, from Wikipedia)~\cite{shao2023character}, or even parametric knowledge in LLMs~\cite{lu-etal-2024-large}, enabling rich persona input. However, abundant personas do not necessarily imply that the personas relevant to a specific context are available. It remains unclear whether models can effectively utilize the most relevant personas from the available information or recognize when crucial personas are not present.


\smallskip\noindent\textbf{Individual simulation without biography} focuses on real-world individuals for applications such as personalized services. Due to privacy constraints and limited access, the persona information available in advance is often sparse. 
Prior work has explored various methods to collect persona information~\cite{Ng2024HowWC,park2024generativeagentsimulations1000,yamashita-etal-2023-realpersonachat}, aiming to gather rich persona of an individual via interviews, pre-specified questions, or questionnaires (\textit{e.g.}, MBTI). However, such methods do not guarantee persona sufficiency in specific contexts. Instead,
focusing on query-relevant personas for each situation provides a more practical approach 
to ensure the persona sufficiency. Identifying relevant personas for each query is thus crucial for improving simulation faithfulness.



In summary, our work focuses on the ill-posed problem in user simulation caused by insufficient relevant personas. Instead of accumulating more persona data, we ask: ``Which persona dimensions are necessary for a given simulation scenario or query?'' 
We formalize this as the task of identifying missing relevant personas in a query-focused context and provide the first benchmark and in-depth analysis for it. By emphasizing minimal yet sufficient persona per query, it follows the ``less is more'' principle, paving the way toward more well-posed, efficient, and faithful simulations.

\subsection{Persona-Augmented Dialogue Dataset}
Existing dialogue datasets with personas, such as PersonaChat~\cite{zhang-etal-2018-personalizing}, Multi-Session Chat~\cite{xu-etal-2022-beyond}, and CharacterEval~\cite{tu-etal-2024-charactereval}, can indeed be used to simulate responses personalized to the target personas. However, these datasets do not provide annotations indicating what personas influence each individual response, making it difficult to study persona sufficiency or to identify missing relevant personas.

In contrast, our PICQ-drama dataset explicitly annotates the missing relevant persona dimensions for each response or decision, enabling controlled evaluation of query-specific persona sufficiency. 
By providing this fine-grained mapping between queries and the personas that shape the responses to them, our dataset 
allows models to not only generate user-like responses but also to identify which persona dimension is still missing, enhancing the fidelity and interpretability of user simulation.

\section{Query-Focused User Simulation} 
\label{sec:task}

A key challenge in studying persona sufficiency is that, in general dialogue, the influence of persona on an utterance is often implicit and difficult to isolate. To make this influence explicit and analyzable, we focus on user responses to questions rather than questions themselves or phatic expressions (\textit{e.g.}, greetings, farewells). Answers to questions often reveal information that directly affects the questioner's subsequent decisions.

To understand the types of questions that naturally arise in dialogue, we conducted a manual analysis of 400 randomly sampled sentences ending with a question mark from the TVShowGuess dataset~\cite{sang2022tvshowguess}. We categorized them into three groups: fact-seeking questions (\textit{e.g.}, ``What is the definition of quantum?'' or ``Where are you from?''), persona-influenced questions, and non-questions. Fact-seeking questions accounted for 61.1\%, persona-influenced questions for 25.6\%, and the remainder were not genuine questions. Although fact-seeking questions constitute the majority, their answers are primarily external facts or personal facts, making them less suitable for studying the influence of persona.
We further examined the persona-influenced questions and found that the majority (approximately 75\%) are choice-based, where the respondent selects from a small set of alternatives, while the rest are open-ended. This observation suggests that choice-based questions are the dominant form of persona-influenced questions in daily dialogue.

Motivated by this data-driven observation, we focus on these persona-influenced choice questions (PICQs), where answers reflect the targets' decisions or opinions shaped by their personas (\textit{e.g.}, ``Would you form a band with me?''). Compared to open-ended questions, PICQs also provide a constrained structure that enables controlled and comparable simulation.

\begin{figure*}[t]
    \centering
    \includegraphics[width=\linewidth]{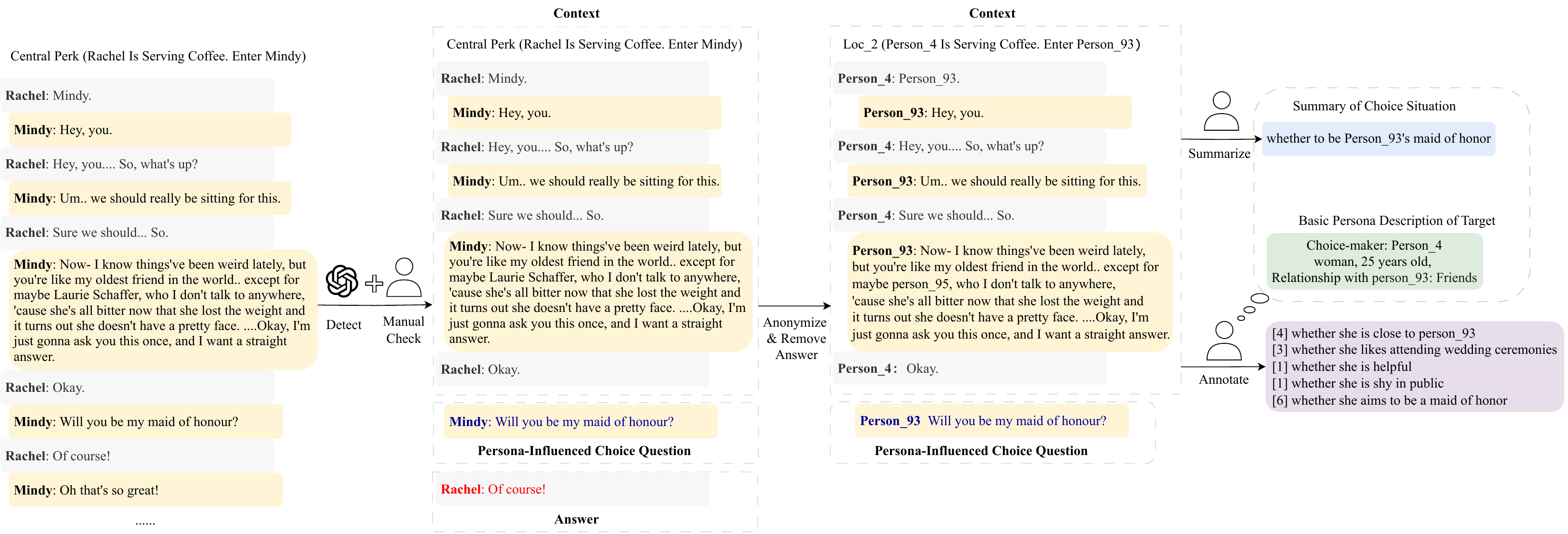}
    \caption{Overview of our approach to acquiring PICQs and annotating the missing relevant personas.}
    \label{fig:overview}
\end{figure*}

In addition to PICQ, we consider including dialogue context preceding each question as the input query. This is because certain questions may appear simple in form (\textit{e.g.}, ``Would you stay here with me?'') but derive their significance from complex preceding situations (\textit{e.g.}, ``they are outside late at night in the rain''). Without context, it would be difficult to accurately interpret the choice being made or simulate a meaningful response. Moreover, some questions are not self-contained (\textit{e.g.}, ``Would you do that with me?'') and cannot be interpreted or answered without the surrounding dialogue.

Following this logic, we provide a basic persona description for the responding character, including gender, age, and their relationship to the questioner (\textit{e.g.}, ``co-worker'' and ``stranger''). These attributes are chosen because they are broadly applicable across diverse questions, commonly adopted in persona-augmented datasets and character simulations~\cite{zhang-etal-2018-personalizing}, and serve as stable anchors for inferring more specific persona dimensions relevant to the choice. Here, a persona dimension is defined as a trait axis whose specific value is currently unknown (\textit{e.g.}, \textit{``whether s/he is shy''}). For brevity, we use persona to refer to the persona dimension in the remainder of the paper.

Therefore, we define the task of identifying missing relevant personas as follows:

\begin{description}
\item{\textbf{Input:}} A dialogue context ($\mathbf{C}$), a PICQ ($\mathbf{Q}$), and a basic persona description ($\mathbf{P}$).
\item{\textbf{Output:}} A set of missing relevant personas ($\mathbf{P_{unk}}$) that are likely to influence the answer ($\mathbf{A}$) to the question as the output.
\end{description}

\section{Data Collection and Annotation}
\label{sec:corpus}
In this section, we describe our dataset derived from the TVshowGuess dataset~\cite{sang2022tvshowguess}, comprising dialogue contexts, persona-influenced choice question (PICQ), a basic persona description, human-identified missing relevant personas, and answers. 
It is constructed from 
the source dialogues in two steps:
i) identifying 
PICQ instances, and ii) annotating the missing relevant personas likely to influence responses, as shown in Figure~\ref{fig:overview}.

\subsection{Source Dialogue Dataset}

We first select a source dialogue dataset that meets two criteria:
i) a realistic setting to support user behavior simulation,
and ii) 
sufficient persona descriptions 
to 
help identify missing relevant personas and assess their impact on the simulation task.

Based on the above two criteria, we selected the TVshowGuess dataset~\cite{sang2022tvshowguess}, which contains English scripts from five popular TV sitcoms. We chose three series, including \textit{Friends}, \textit{Frasier}, and \textit{The Office}, balancing topical diversity with annotation feasibility. These series cover themes like friendship, romance, family, and 
career, aligning well with our focus on everyday choice-making. We use the first three seasons of them as the source dialogues.

\subsection{Discovering PICQs and Answers}


Our first stage of annotation is to identify PICQs and their 
answers from the source dialogues. The definition of PICQs is discussed in \S~\ref{sec:task}. A corresponding answer to a PICQ is defined as the immediate next utterance following PICQ that clearly chooses one of the alternatives implied or listed by the question. Restricting the answer to the next utterance ensures that it reflects the respondent's initial persona-based intent and avoids incorporating later choices that may result from discussions. 


To reduce annotation cost, we first prompt GPT-4.1 to detect potential PICQs and their 
answers. Three human annotators (the first, second, and third authors) then verify whether each candidate pair satisfies our task criteria. Refer to Appendix~\ref{prompt_identify} for the prompts and Appendix~\ref{ins_anno} for the annotation guidelines. Each annotator reviews two-thirds of the data to ensure overlap and allow measurement of inter-annotator agreement. The average Cohen's $\kappa$ between annotator pairs is 0.740, indicating substantial agreement. Disagreements are then resolved by discussion to ensure data quality. 

\subsection{Annotating Missing Relevant Personas}
\label{subsec:anno_un}
Our second stage of annotation is to identify missing relevant personas for each PICQ, given its context, and the respondent's basic persona (\S~\ref{sec:task}). We first define what missing personas can be annotated, and then introduce our annotation process.

Based on the persona definitions from previous work~\cite{chuang-etal-2024-beyond,yuan-etal-2024-evaluating}, we formulate seven categories of personas: personality, beliefs, tastes, relationship, attributes, goals, and experience. See category details \S~\ref{sec:category} and annotation examples in \S~\ref{example}.
Coarse-grained categories alone are insufficient to capture specific, missing relevant persona descriptions. However, allowing fully free-form text makes it hard to determine whether two descriptions refer to the same underlying persona. To balance structure and expressiveness, we developed ten lexico-syntactic templates per category (\textit{e.g.}, ``\textit{whether s/he (dis)likes VP},'' where VP stands for a verb phrase), based on patterns observed in preliminary annotations. Refer to the full list of templates in Appendix~\ref{ins_anno}.
We then describe the process of annotating missing relevant persona descriptions, which comprises three steps: (1) anonymization, (2) query-focused summarization, and (3) persona annotation.

\paragraph{Anonymization}
To ensure that persona identification relies solely on the provided basic persona description rather than human annotators' prior knowledge or models' parametric knowledge, 
we anonymize the dialogue data. Specifically, we replace all character names, organizations, geopolitical entities, facilities, and locations with placeholders (\textit{e.g.}, ``\textit{$person_1$},'' ``\textit{$org_1$}'') using a named entity recognition (NER) following~\citet{sang2022tvshowguess}.

\paragraph{Query-Focused Summarization}
Next, we produce a self-contained summary that clarifies the choice-making situation, ensuring that subsequent persona annotations are grounded in the same interpretation of the context. We ask three annotators to summarize each PICQ along with its preceding dialogue context in one sentence, as one example shown in Figure~\ref{fig:task}, and the instruction is shown in Appendix~\ref{ins_anno}. Each annotator works on two-thirds of the data, enabling overlap and cross-validation. The average agreement rate between annotator pairs is 88\%, and consistency is judged by a natural language inference (NLI) model\footnote{\url{https://huggingface.co/cross-encoder/nli-deberta-v3-base}\label{fn:nli}}. Disagreements are resolved through discussion. 

\paragraph{Missing Relevant Persona Annotation}
Given the PICQ, context, summary, and the basic persona description, annotators are instructed to identify up to five missing persona descriptions most likely to influence the answer. For each, they first select a persona category from our predefined set (\textit{e.g.}, Goal) and then describe the persona using a specific linguistic pattern associated with that category (Refer to Appendix~\ref{ins_anno} for category details). Multiple personas per category are allowed, and irrelevant categories may be skipped.
Annotations should prioritize personas serving as strong motivations, necessary conditions, and critical factors behind the choice when there are more than five relevant personas. Annotators are encouraged to generalize personas without losing core meaning and to avoid overly specific phrasing (\textit{e.g.}, ``\textit{whether he likes Indian chicken curry}'' to ``\textit{whether he likes curry}'').

\begin{table}[t]

  \centering
  \tabcolsep 4pt
\small 
  \begin{tabular}{lr}
    \toprule
    Dialogue scenes w/ PICQ-answer pairs  & 289  \\
     PICQ-answer pairs influenced by persona & 300\\
     Total number of characters as listeners& 60 \\
     \midrule
     Ave. number of utterances in dialogue context &  22.69 \\
    Ave. number of tokens per utterance & 17.23  \\
    \midrule
     Ave.\ identified relevant personas (after merging) & 3.53  \\
     Ave. number of tokens per relevant persona &  6.58\\
    \bottomrule
  \end{tabular}
  \caption{Dataset PICQ-drama: statistics.}
  \label{tab:st}
\end{table}

Each annotator labels two-thirds of the data to ensure overlap. Given that identifying relevant personas involves subjective judgment, prioritization, and potentially incomplete enumeration of personas, this overlap is designed to assess inter-annotator agreement and ensure annotation reliability. We automatically evaluate persona alignment using category matching and an NLI model.\footref{fn:nli} Persona descriptions are considered to refer to the same persona if the NLI model predicts either entailment or contradiction (\textit{e.g.}, ``\textit{whether he is introverted}'' and ``\textit{whether he is extroverted}'' are treated as aligned). On average, 59\% of each annotator's persona annotations overlap with those of others.\footnote{We recruit another external annotator to confirm the solidness of our annotation. Reading only the instructions, the annotator achieved 50\% overlap with 
the annotators on 10\% of the instances, demonstrating the task's reproducibility.} The non-overlapping cases are partly attributed to the subjective nature of the task and the annotators' differing background knowledge and lived experiences, which may influence what aspects of the persona they perceive as relevant (\textit{e.g.}, for a spending-related decision, annotators with different economic backgrounds may disagree on whether ``\textit{whether his financial status is good}'' is relevant).

For the final dataset, we prioritize personas that are annotated by multiple annotators, keeping only one instance for each agreed-upon persona. If fewer than five such agreed-upon personas are available, we supplement them with additional non-overlapping ones. If more than five candidate personas exist, the annotators select the five most salient ones through discussion and reconciliation. Refer to Appendix~\ref{ins_anno} for the annotation instructions. The dataset statistics are shown in Table~\ref{tab:st}.

\section{Evaluation of Identifying missing Relevant Persona}
\label{sec:intrinsic}
To empirically evaluate the task of identifying missing relevant personas, we conduct a series of experiments using LLMs (\S~\ref{sec:models}) to identify missing relevant persona and evaluating the indentified persona via various metrics (\S~\ref{sec:metrics}). Specifically, we answer the three research questions; 
RQ1: How well can leading LLMs identify missing relevant personas from context? (\S~\ref{mainRe});
RQ2: What cognitive patterns or differences emerge in their performance, especially compared to humans? (\S~\ref{cognitive});
RQ3: Can effective instruction strategies enhance model performance on this task? (\S~\ref{ablation}).

\subsection{Models}\label{sec:models}
We evaluate closed- and open-source LLMs covering a range of architectures and scales. We used GPT-4.1-2025-04-14 as a strong closed-source LLM. We hereafter refer to it as GPT-4.1. We used Llama3.1-8B-Instruction,\footnote{\url{https://huggingface.co/meta-llama/Llama-3.1-8B-Instruct}} Llama3.1-70B-Instruction,\footnote{\url{https://huggingface.co/meta-llama/Llama-3.1-70B-Instruct}} Qwen3-8B,\footnote{\url{https://huggingface.co/Qwen/Qwen3-8B}} and Qwen3-32B\footnote{\url{https://huggingface.co/Qwen/Qwen3-32B}} models as open-source LLMs.

    
\paragraph{Instruction Strategies}
We test two types of instructions to solve the task, including one proposed for this task. 
\textbf{Baseline Prompting} directly asks the model to perform the identification task. 
In contrast, our proposed \textbf{Multi-task Prompting} first instructs the model to summarize the choice situation before identifying the personas, aiming to improve the comprehension, and then instructs the model to generalize the personas after the identification, aiming to avoid over-specific outputs. Refer to the prompts in  Appendix~\ref{prompt_identify}.

\paragraph{Human (Oracle)} The annotations in our PICQ-drama dataset, listed and merged by the two human annotators for each query, serve as the ground truth for comparison across all metrics.

\begin{table*}[t]
  \small
  \centering
  \begin{tabular}{lcccccc}
    \toprule
    \multirow{2}{*}[-2pt]{\textbf{Models}} &  \multirow{2}{*}[-2pt]{\textbf{Influence}}  & \multirow{2}{*}[-2pt]{\textbf{Inaccessibility $\downarrow$}} & \multicolumn{3}{c}{\textbf{Fidelity}} & \multirow{2}{*}[-2pt]{\textbf{Ave.\ Per.}} \\
    \cmidrule(lr){4-6}
     &    & & \textbf{Precision} & \textbf{Recall} & \textbf{\textit{F}}$_\mathbf{1}$ & \\
    \midrule
     Llama3.1-8B & $1.238$ & $\mathbf{1.385}$& $0.318$ & $0.536$ &$0.399$& $5.00$\\ 
     Llama3.1-8B-Multi & $1.397$ & $1.395$& $0.390$ & $0.450$ &$0.418$& $4.06$\\
     Llama3.1-70B & $1.525$ & $1.504$& $0.271$ & $0.359$ &$0.309$& $4.65$\\ 
     Llama3.1-70B-Multi & $1.621$ & $1.430$& $0.318$ & $0.380$ &$0.346$& $4.22$\\
     Qwen3-8B & $1.377$ & $1.485$& $0.345$ & $0.456$ &$0.393$& $4.66$\\ 
     Qwen3-8B-Multi & $1.567$ & $1.608$ & $0.350$ & $0.442$ &$0.391$&$4.45$\\ 
     Qwen3-32B & $1.510$ & $1.558$ & $0.399$ & $0.567$ &$0.468$&$5.00$\\ 
     Qwen3-32B-Multi & $1.589$ & $1.513$ & $\mathbf{0.409}$ & $\mathbf{0.581}$ &$\mathbf{0.480}$&$4.99$\\ 
    \midrule
     GPT-4.1 & $1.711$ & $1.540$& $0.333$ & $0.425$ &$0.374$& $4.49$\\ 
     GPT-4.1-Multi & $\mathbf{1.772}$ & $1.571$& $0.306$ & $0.294$ &$0.300$& $3.38$\\ 
     \midrule
     Human (Oracle) & $1.648$ & $1.394$& $-$ & $-$ &$-$& $3.53$\\ 
    \bottomrule
  \end{tabular}
  \caption{Results of identifying missing relevant personas.}
  \label{tab:inResult}
\end{table*}


\subsection{Metrics}\label{sec:metrics}
We design a multi-faceted evaluation scheme to provide an assessment of model performance: 

\smallskip\noindent\textbf{Influence} measures the perceived impact of a missing persona on a character's decision. We use
a 3-point Likert scale (0: irrelevant, 1: minor, 2: key).
This metric reflects a model's ability to provide in-depth insights. Based on the manually summarized choice situation (\S~\ref{subsec:anno_un}), each persona is scored on whether it is irrelevant to the choice, slightly shapes preferences, or serves as a central motivation or constraint. We use GPT-4.1 (Temperature = 0.7) for full-scale automatic scoring. To validate this approach, we compared its ratings on 100 samples against human annotations, achieving a Cohen's $\kappa$ of 0.658. The human inter-annotator agreement on the same set reached a $\kappa$ of 0.831. Refer to Appendix~\ref{prompt_judge} for detailed definitions and the prompt.

\smallskip\noindent\textbf{Inaccessibility} measures the difficulty of acquiring a missing persona, which also serves as a proxy for its privacy level. We use a 3-point Likert scale (0: Very Easy, 1: Easy, 2: Hard), where scores reflect whether a persona is observable by strangers (\textit{e.g.}, gender), known to acquaintances, or requires close friendship. We use GPT-4.1 for full-scale automatic scoring. To validate this method, we compared its ratings on 100 samples with human annotations, achieving a Cohen's $\kappa$ of 0.501. The inter-annotator agreement between two humans on the same set was a substantial $\kappa$ of 0.618. Refer to Appendix~\ref{prompt_judge} for the detailed prompt.


\smallskip\noindent\textbf{Fidelity} 
measures the semantic alignment between model-generated personas and the human-annotated gold references. The primary challenge is that free-text descriptions can be lexically different yet refer to the same underlying semantic factor. To handle this, we employ a two-step matching criterion: two personas are considered a match if (1) they belong to the same predefined persona category (\textit{e.g.}, Beliefs), and (2) both descriptions probe the same persona with NLI (\S~\ref{subsec:anno_un}), persona descriptions are considered to refer to the same persona if the NLI model predicts either entailment or contradiction (\textit{e.g.}, ``whether he likes X'' and ``whether he dislikes X''). Based on this matching logic, we compute standard Precision, Recall, and $F_1$ scores to quantify the fidelity of a model's output.

\smallskip\noindent\textbf{Average number of missing relevant personas (Ave.\ Per.)} is reported to show how many pieces of missing relevant personas the model identifies; we calculate the average number of missing relevant personas generated per PICQ of the models.

We perform paired bootstrap significance testing ($\alpha=0.05$) on Influence, Inaccessibility, and $F_1$ scores to ensure that the claims in~\S~\ref{mainRe} and~\S~\ref{ablation} are statistically significant.

\subsection{Main Results}
\label{mainRe}
Table~\ref{tab:inResult} shows the main results of our experiments. A key observation is that no single model excels across all metrics. Instead, the results reveal a complex, scale-dependent relationship between a model's ability to imitate human patterns (Fidelity) and its ability to generate profound analytical explanations (Influence). 

Focusing on the fidelity dimension, as measured by the $F_1$ score, Qwen3-32B and Qwen3-32B-Multi achieve the highest Recall and $F_1$ scores, indicating that their generated personas align most closely with our human-annotated ground truth.
In contrast, when evaluating for influence, as measured by the Influence score, GPT-4.1-Multi stands out, achieving the top score of 1.772. Most notably, this score surpasses even the Human (Oracle) baseline of 1.648. This suggests that GPT-4.1, when guided by our multi-task prompt, can identify personas perceived as even more impactful or fundamental to the decision than those articulated by humans.

Observing the scaling dynamics of fidelity and influence, we find that fidelity follows an inverted U-shaped trend. As model size increases from small (\textit{e.g.}, Llama3.1-8B) to medium (\textit{e.g.}, Qwen3-32B), fidelity improves. However, it declines for the largest models (e.g., Llama3.1-70B).
In contrast, influence consistently increases with model size. Larger models are better at inferring missing personas, but as models become sufficiently large, their predictions become less aligned with human judgment patterns.

The Inaccessibility metric provides another critical layer of analysis. The Human (Oracle) exhibits the near lowest inaccessibility score (1.394), demonstrating remarkable efficiency. This aligns with the principle of ``cognitive economy'' in psychology~\cite{rescher2017cognitive}: humans are exceptionally skilled at identifying highly accessible (i.e., low inaccessibility) personas that still possess strong explanatory power (i.e., high influence). While some models like Llama3.1-8B achieve low inaccessibility, they do so at the cost of significantly lower influence.

Finally, the impact of our multi-task instruction strategy is consistent across the board. It systematically boosts the influence score for every model family, while also consistently reducing the Average Personas (Ave. Per.) generated. This confirms its role in encouraging models to perform analytical synthesis rather than simple enumeration. Some generated examples are shown in Appendix~\ref{example}.

\begin{table*}[t]
\small
\centering
\begin{tabular}{lccccccc}
\toprule
\textbf{Model} & \textbf{Personality} & \textbf{Belief} & \textbf{Taste} & \textbf{Relationship} & \textbf{Attribute} & \textbf{Goal} & \textbf{Experience} \\
\midrule
Llama3.1-8B & $\mathbf{28.9}$ & $11.4$ & $18.1$ & $\phantom{0}3.4$ & $\mathbf{20.6}$ & $11.0$ & $\phantom{0}6.6$ \\
Llama3.1-70B & $19.1$ & $\phantom{0}1.2$ & $15.2$ & $\phantom{0}2.8$ & $\mathbf{40.3}$ & $\mathbf{21.0}$ & $\phantom{0}0.2$ \\
Qwen3-8B & $\mathbf{21.3}$ & $17.7$ & $\mathbf{22.9}$ & $\phantom{0}5.8$ & $14.0$ & $18.1$ & $\phantom{0}0.3$ \\
Qwen3-32B & $\mathbf{26.9}$ & $10.5$ & $19.2$ & $14.2$ & $12.9$ & $14.9$ & $\phantom{0}1.3$ \\
GPT-4.1 & $16.8$ & $\mathbf{21.4}$ & $\mathbf{27.9}$ & $\phantom{0}5.1$ & $12.1$ & $11.2$ & $\phantom{0}5.5$ \\
Human (Oracle) & $\mathbf{28.9}$ & $14.3$ & $19.3$ & $\mathbf{25.2}$ & $\phantom{0}6.0$ & $\phantom{0}4.2$ & $\phantom{0}2.3$ \\
\bottomrule
\end{tabular}
\caption{Distribution of Identified Relevant Persona Types (\%). (percentages higher than 20\% are bold).}
\label{tab:distribution}
\end{table*}

\subsection{Analysis of Cognitive Differences}
\label{cognitive}
Our results not only quantify model performance but also provide a window into the distinct cognitive models of different intelligent agents. We first diagnose the fundamental differences in attribution patterns between humans and LLMs. We then uncover the unique cognitive strengths of each agent type. Finally, based on these insights, we propose a synergistic framework that leverages these complementary advantages.

Table~\ref{tab:distribution} shows the persona category distributions, revealing that humans and LLMs operate with fundamentally different cognitive patterns. Human annotators exhibit a clear cognitive model grounded in social context, heavily favoring Personality (28.9\%), Taste (19.3\%), and Relationship (25.2\%). 
The most significant difference between humans and LLMs lies in two categories: humans prioritize Relationship, while LLMs, as a group, consistently favor Goal. This points to a core divergence: human reasoning is deeply embedded in social dynamics, whereas LLMs appear to operate under a more utilitarian, task-oriented framework, assuming a goal-driven motive behind actions.
While LLMs share this general tendency, there are important outliers. Qwen3-32B, with its relatively high emphasis on Relationship, stands out as the most ``human-like'' LLM, explaining its top-tier fidelity. In contrast, the Llama series displays a unique bias towards Attribute, offering a different perspective on choice-making.

\begin{figure}[t]
\centering
\includegraphics[width=\linewidth]{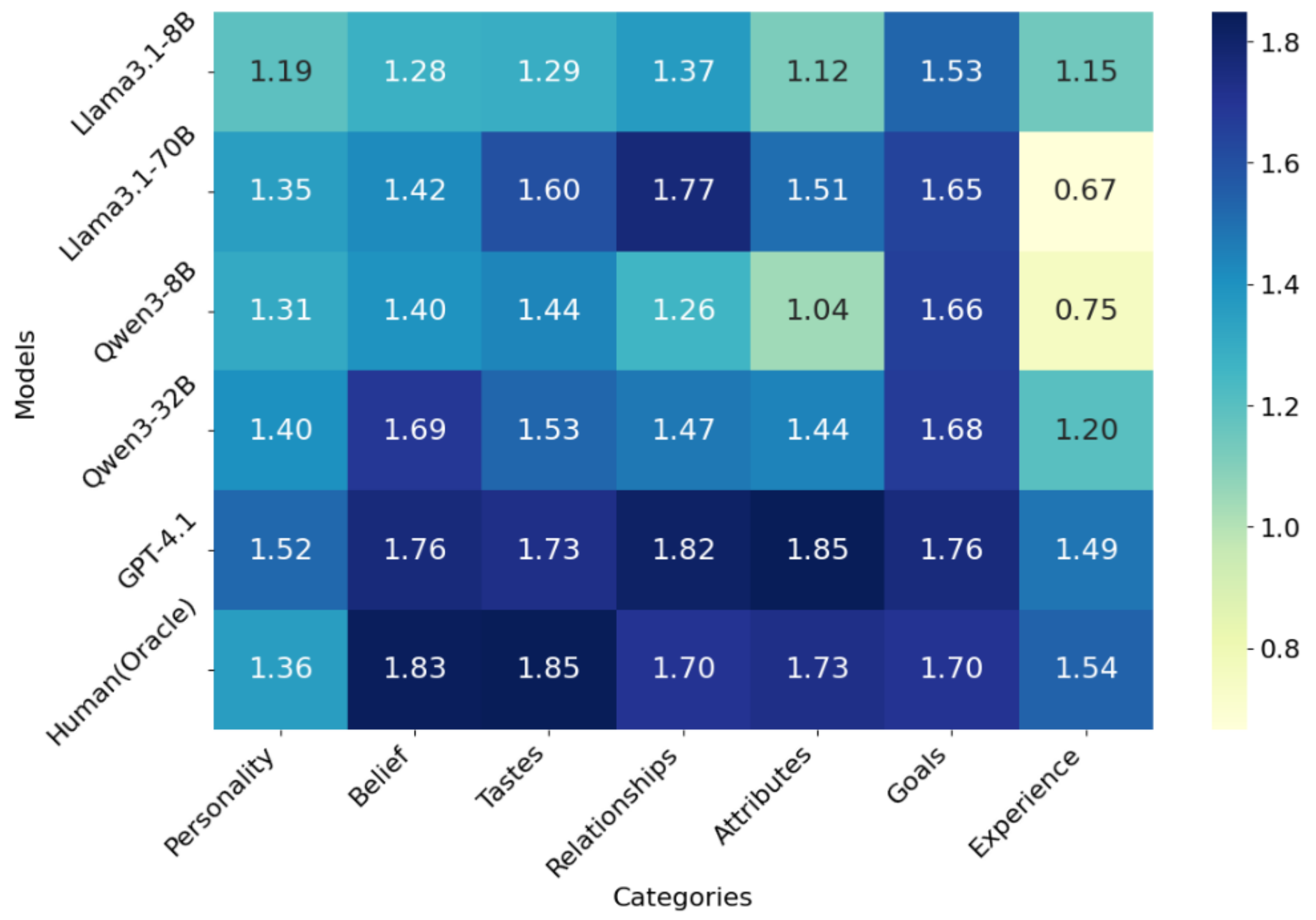}   
\caption{Influence heatmap of identified personas according to categories.}
\label{fig:inf}
\end{figure}

\begin{figure}[t]
\centering
\includegraphics[width=\linewidth]{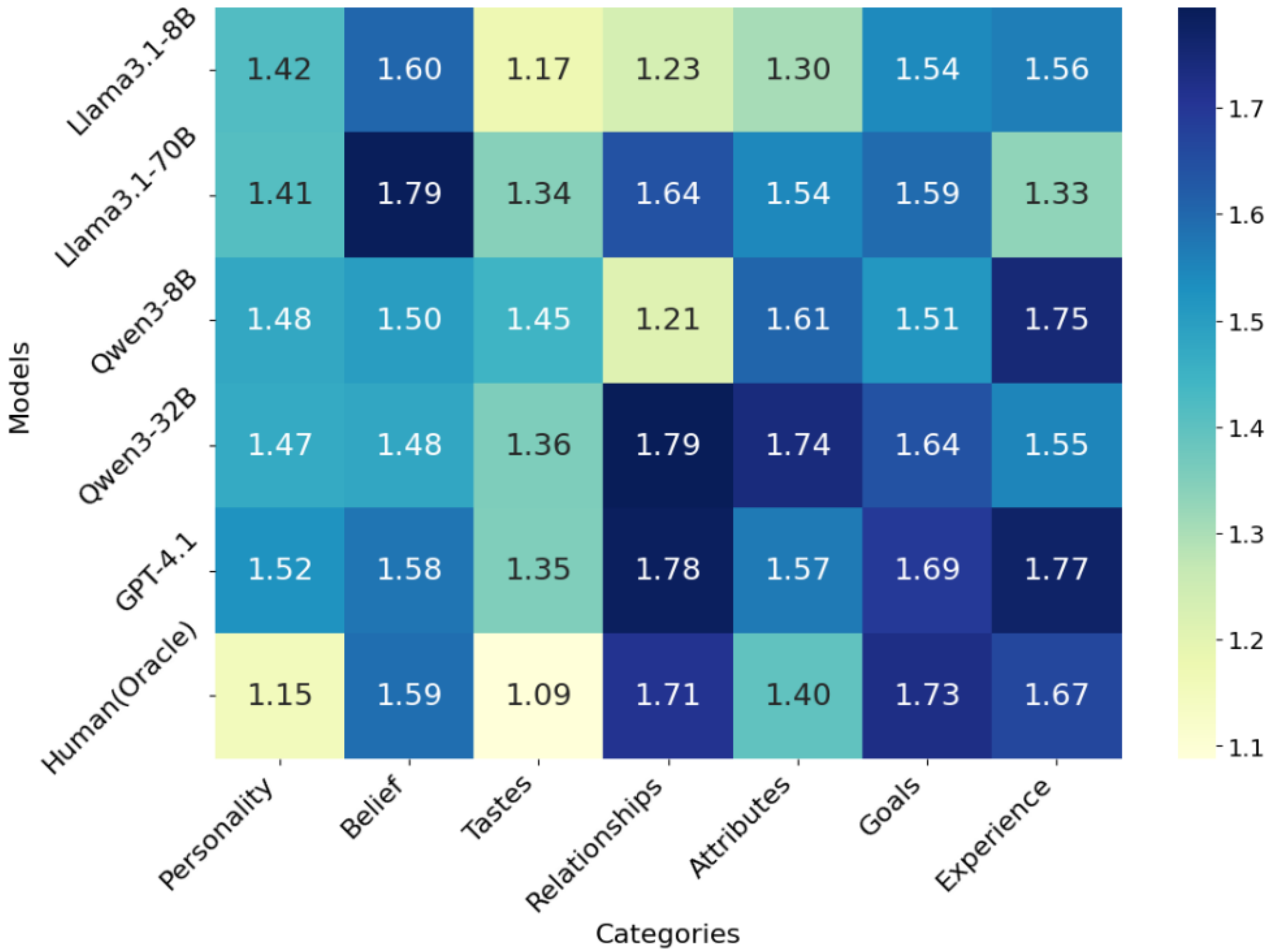}   
\caption{Inaccessibility heatmap of identified personas according to categories.}
\label{fig:ina}
\end{figure}

Figures~\ref{fig:inf} and~\ref{fig:ina} allow us to move to identifying the characteristics of each persona category and the unique strengths of each agent. The categories themselves exhibit distinct profiles. Personality acts as a global trait with moderate influence and low accessibility. Belief, Goal, and Relationship function as deep motivators, being both high-impact and hard to acquire. Taste serves as a direct driver, being highly influential and easily accessible. 
We can observe that the GPT-4.1 model achieves consistently high Influence across nearly all categories, capable of uncovering high-impact motives. While humans excel at identifying personas in the Personality, Taste, and Attribute categories that yield extremely high influence for a very low inaccessibility cost. This aligns with the principle of ``cognitive economy,'' positioning humans as experts in finding high-yield, low-effort explanations. 

Our analysis suggests that the path forward is not selecting a single best agent, but combining complementary persona sources into a synergistic framework for identifying missing relevant personas.
For dataset construction, our task can guide developers to collect minimal yet sufficient personas, improving simulation fidelity. This can be implemented as a three-stage process: divergent persona generation, convergent filtering to remove redundancy, and final human selection to ensure relevance.
For simulating a specific individual, the framework can operate as a persona completion process, where a model identifies missing persona dimensions for the current scenario and prompts the user (or another module) to provide them.

\begin{table*}[t]
  \small
  \centering
  \begin{tabular}{lcccccc}
    \toprule
    \multirow{2}{*}[-2pt]{\textbf{Models}} &  \multirow{2}{*}[-2pt]{\textbf{Influence}}  & \multirow{2}{*}[-2pt]{\textbf{Inaccessibility $\downarrow$}} & \multicolumn{3}{c}{\textbf{Fidelity}} & \multirow{2}{*}[-2pt]{\textbf{Ave.\ Per.}} \\
    \cmidrule(lr){4-6}
     &    & & \textbf{Precision} & \textbf{Recall} &\textbf{\textit{F}}$_1$ & \\
    \midrule
     Qwen3-32B-Multi & $1.589$ & $\mathbf{1.513}$ & $0.409$ & $0.581$ &$0.480$&$4.99$\\ 
     $-$ Summarization & $1.500$ & $1.541$& $0.411$ & $0.564$ &$0.476$& $4.83$\\
     $-$ Generalization & $\mathbf{1.594}$ & $1.556$& $0.410$ & $0.590$ &$\mathbf{0.484}$& $5.00$\\ 
     $-$ Summarization \& Generalization & $1.510$ & $1.558$ & $0.399$ & $0.567$ &$0.468$&$5.00$\\
     \midrule
     GPT-4.1-Multi & $1.772$ & $1.571$& $0.306$ & $0.294$ &$0.300$& $3.38$\\
     $-$ Summarization & $\mathbf{1.794}$ & $\mathbf{1.538}$& $0.347$ & $0.274$ &$0.306$& $2.78$\\ 
     $-$ Generalization & $1.698$ & $1.556$ & $0.316$ & $0.430$ &$0.364$&$4.79$\\
     $-$ Summarization \& Generalization & $1.711$ & $1.540$& $0.333$ & $0.425$ &$\mathbf{0.374}$& $4.49$\\
    \bottomrule
  \end{tabular}
  \caption{Ablation results of our multi-task instruction strategy.}
  \label{tab:ablation}
\end{table*}

\subsection{Ablation Study}
\label{ablation}
To understand the contribution of each component in our multi-task instruction strategy, we conduct an ablation study. Our strategy combines two steps: summarizing the choice situation and generalizing the identified personas. We analyze the impact of these components on two models that have the best performance on influence and fidelity aspects: GPT-4.1 and Qwen3-32B. 

Table~\ref{tab:ablation} shows the results.
The Summarize component yields a significant improvement only for Qwen3-32B in terms of Influence. This supports its intended role of helping the model correctly interpret the specific choice situation. For GPT-4.1, no significant gain is observed, suggesting its baseline comprehension is already sufficient.
In contrast, the Generalize component has a significant effect only on GPT-4.1, increasing influence while reducing fidelity. However, the resulting abstraction exceeds the level of cognitive effort typically exercised by human annotators, leading to aggressive merging of persona dimensions, fewer generated personas, and lower fidelity to human patterns.
Overall, the Summarize component benefits mid-sized models by improving the understanding of the choice situation, while the Generalize component primarily affects large models by triggering deeper abstraction, increasing influence at the cost of fidelity.

\section{Conclusions}
This work highlights a common oversight in user simulation: the assumption that provided persona information is sufficient. We formalize this as a new task, identifying missing relevant persona dimensions, and present the first benchmark for its evaluation.
Using our PICQ-drama dataset, we demonstrate the feasibility of applying LLMs to this task. Our results show that the ability to detect influential missing personas generally increases with model scale. The discovery of an inverted U-shaped fidelity curve, linked to the concept of human ``cognitive economy,'' offers a novel lens for comparing human and LLM cognition. Our further analysis shows the cognitive differences within LLMs, as well as between LLMs and humans. Besides, we design a multi-task instruction strategy that improves the LLMs' ability to identify missing personas that better influence the choices. 

Future work will focus on collecting the identified missing personas to evaluate their direct impact on the downstream simulation tasks.

\section*{Acknowledgement}
This work was supported by Institute for Digital Observatory, the University of Tokyo.

\section*{Limitations}
Our study is limited to English-language data. Differences in language and sociocultural background, whether in model training or human annotation, may lead to divergent interpretations of what constitutes a relevant persona. As such, the identified personas and their perceived influence on user responses may vary across linguistic and cultural contexts, suggesting that further exploration is needed to understand and generalize these findings across languages and cultures.

Our study is based on drama scripts rather than spontaneous, real-world conversations. This choice was made primarily to navigate the significant ethical challenges, such as privacy and consent, associated with collecting and analyzing authentic personal dialogues. To maximize the resemblance to reality, we 
selected scripts from sitcoms and dramas that focus on everyday life, interpersonal relationships, and common choice-making scenarios. However, an unavoidable gap remains. Scripted dialogue is typically more structured, coherent, and sometimes theatrically heightened compared to authentic speech, which is often disfluent and fragmented. Future research should aim to validate and extend our findings on datasets of anonymized, ethically-sourced real-world conversations to assess the generalizability of our findings.

\bibliography{custom}

\appendix

\section{Appendix}
\label{sec:appendix}

\subsection{Categories of Personas}
\label{sec:category}
\begin{itemize}
    \item \textbf{Personality} Stable psychological traits such as extroversion, or emotional sensitivity. These traits influence how a person tends to behave or react in various situations.
    \item \textbf{Beliefs} Enduring convictions or values, including moral principles, or political stances, shaping a person's judgment of what is right.
    \item \textbf{Tastes} Personal preferences for things like food, music, or entertainment. Tastes can strongly affect choices involving consumption, participation, or lifestyle.
    \item \textbf{Relationship} Social ties and interpersonal history with other characters, such as being a friend, sibling, or coworker. These relationships influence the level of trust, obligation, or emotional support, which can significantly shape one's choice.
    \item \textbf{Attributes} Basic biographical or demographic characteristics, such as income level, occupation, or cultural background. These factors may constrain or inform choices due to physical capability or role expectations
    \item \textbf{Goals} Current intentions, needs, or objectives a person is trying to achieve. Goals directly impact choices by framing what is desirable or prioritized in a given situation.
    \item \textbf{Experience} Experience would be chosen only if the specific past event memory itself is directly influencing the choice, rather than having been internalized in other categories.
\end{itemize}

\subsection{Generated Examples}
\label{example}
Tables~\ref{tab:examp_1} and~\ref{tab:examp_2} some generated examples, we picked GPT-4.1-Multi (high influence) and Qwen3-32-Multi (high fidelity) settings, and human annotations. These examples highlight the different tendencies of models. GPT-4.1-Multi tends to generate more concise and impactful explanations, whereas Qwen3-32B-Multi produces more constrained, context-grounded personas.

\subsection{Instructions for Annotation}
\label{ins_anno}
Table~\ref{tab:step1_human} shows the instructions for human annotators to check potential PICQs and their corresponding answers. Table~\ref{tab:step2_summarization} shows the instructions for human annotators to write query-focused summarization. Table~\ref{tab:step2_identify} shows the instructions for human annotators to annotate the missing relevant personas

\subsection{Prompts for Identifying PICQs and missing Relevant Personas}
\label{prompt_identify}
Table~\ref{tab:step1_gpt} shows the prompts for LLMs to detect potential PICQs and their corresponding answers. Table~\ref{tab:prompt_identify} shows the prompts for LLMs to identify the missing relevant personas. Table~\ref{tab:prompt_identifycot} shows the prompts for LLMs to identify the missing relevant personas using a multi-task instruction strategy.

\subsection{Prompts for LLM-as-Judge}
\label{prompt_judge}
Table~\ref{tab:prompt_Influence} shows the prompts for LLMs to evaluate the Influence score of the identified missing relevant personas. Table~\ref{tab:prompt_Difficulty} shows the prompts for LLMs to evaluate the inaccessibility score of the identified missing relevant personas.

\begin{table*}[t]
  \footnotesize
  \begin{tabular}{p{0.97\linewidth}}
    \toprule
    \textbf{[Example 1] Context \& Identified Missing Relevant Persona Dimensions} \\
    \midrule
    \textbf{Context:} Background Scene 6: Central Perk (Rachel Is Serving Coffee. Enter Mindy) \\
    (2) Rachel: Mindy. \\
    (3) Mindy: Hey, you. \\
    (4) Rachel: Hey, you.... So, what's up? \\
    (5) Mindy: Um.. we should really be sitting for this. \\
    (6) Rachel: Sure we should... So. \\
    (7) Mindy: Now- I know things've been weird lately, but you're like my oldest friend in the world.. except for maybe Laurie Schaffer, who I don't talk to anywhere, 'cause she's all bitter now that she lost the weight and it turns out she doesn't have a pretty face. ....Okay, I'm just gonna ask you this once, and I want a straight answer. \\
    (8) Rachel: Okay. \\
    (9) Mindy: \textbf{Will you be my maid of honor?} \\[1em]

    \textbf{GPT-4.1-Multi} \\
    \hspace{1em} $\bullet$ \textbf{tastes}: whether she likes or dislikes being involved in major ceremonial roles such as maid of honour \\
    \hspace{1em} $\bullet$ \textbf{personality}: whether she is outgoing and enjoys social responsibility \\
    \hspace{1em} $\bullet$ \textbf{relationships}: whether she feels emotionally connected to Mindy at the present time \\
    \hspace{1em} $\bullet$ \textbf{experience}: whether she has had past experiences that shape her willingness to accept this role \\
    \hspace{1em} $\bullet$ \textbf{attributes}: whether she has enough free time or flexibility in her personal schedule to take this on \\[0.5em]

    \textbf{Qwen3-32B-Multi} \\
    \hspace{1em} $\bullet$ \textbf{beliefs}: whether she believes it is important to support close friends in significant life events \\
    \hspace{1em} $\bullet$ \textbf{personality}: whether she is reliable \\
    \hspace{1em} $\bullet$ \textbf{relationships}: whether she is loyal to Mindy \\
    \hspace{1em} $\bullet$ \textbf{goals}: whether she aims to maintain a strong friendship with Mindy \\
    \hspace{1em} $\bullet$ \textbf{personality}: whether she is comfortable with taking on responsibilities \\[0.5em]

    \textbf{Human} \\
    \hspace{1em} $\bullet$ \textbf{relationships}: whether she is close to Mindy \\
    \hspace{1em} $\bullet$ \textbf{tastes}: whether she likes attending wedding ceremonies \\
    \hspace{1em} $\bullet$ \textbf{personality}: whether she is helpful \\
    \hspace{1em} $\bullet$ \textbf{personality}: whether she is shy in public \\
    \hspace{1em} $\bullet$ \textbf{goals}: whether she aims to be a maid of honor \\
    
    \bottomrule
  \end{tabular}
  \caption{Generated example 1.}
  \label{tab:examp_1}
\end{table*}

\begin{table*}[t]
  \footnotesize
  \begin{tabular}{p{0.97\linewidth}}
    \toprule
    \textbf{[Example 2] Context \& Identified Missing Relevant Persona Dimensions} \\
    \midrule
    \textbf{Context:} Background ([SCENE\_BREAK]) \\
    (1) Michael: I should have never let the Temp touch this thing. I had all these great icons and now I have four folders. So.. \\
    (2) Dwight: It's actually better this way. \\
    (3) Michael: No it's not. Because I could just click on the icon and then I'm onto--- \\
    (4) Dwight: Michael, could I ask you something? I wanted to ask your permission to ask out Katy. I know it's against the rules and everything. Because... \\
    (5) Michael: No, no, no it's not against the rules. She's not a permanent employee so it's not. \\
    (6) Dwight: Thank you, Michael. I appreciate this so much. \\
    (7) Michael: But I think you should just know that I am going to be giving her a ride home later. \\
    (8) Dwight: What? \\
    (9) Michael: She asked me for a ride and so I am going to give her a ride home. \\
    (10) Dwight: Is that all it is? Just a ride home? Like a taxicab? \\
    (11) Michael: Well, might be a ride home. Might be a ride home and we stop for coffee and dot-dot-dot... \\
    (12) Dwight: Please. Please, I am your inferior and I'm asking you this favor. \textbf{Can you promise me that it will just be a ride home?} \\[1em]

    \textbf{GPT-4.1-Multi} \\
    \hspace{1em} $\bullet$ \textbf{beliefs}: whether he values professional ethics in interpersonal relationships \\
    \hspace{1em} $\bullet$ \textbf{personality}: whether he is empathetic or enjoys holding power in ambiguous situations \\[0.5em]

    \textbf{Qwen3-32B-Multi} \\
    \hspace{1em} $\bullet$ \textbf{personality}: whether he is trustworthy \\
    \hspace{1em} $\bullet$ \textbf{personality}: whether he is considerate \\
    \hspace{1em} $\bullet$ \textbf{beliefs}: whether he believes it’s important to respect others’ feelings \\
    \hspace{1em} $\bullet$ \textbf{relationships}: whether he is friendly with Katy \\
    \hspace{1em} $\bullet$ \textbf{goals}: whether he aims to maintain a professional relationship with Dwight \\[0.5em]

    \textbf{Human} \\
    \hspace{1em} $\bullet$ \textbf{relationship}: whether he is fond of Katy \\
    \hspace{1em} $\bullet$ \textbf{relationship}: whether he is close with Dwight \\
    \hspace{1em} $\bullet$ \textbf{beliefs}: whether he believes it's fine to get an employee to be his girlfriend \\
    \hspace{1em} $\bullet$ \textbf{attributes}: whether his pattern status is 'in a relationship' \\
    \hspace{1em} $\bullet$ \textbf{personality}: whether he is gentle \\
    
    \bottomrule
  \end{tabular}
  \caption{Generated example 2.}
  \label{tab:examp_2}
\end{table*}

\begin{table*}[t] 
  \footnotesize
  \begin{tabular}{p{0.97\linewidth}} 
    \toprule
    \texttt{Please review the given dialogue and identify all specific question-answering pairs that consist of:} \\
        \texttt{\hspace{1em}- An utterance that asks a specific listener to choose among specific options whose choice will implicitly depend on their personas (\textit{e.g.}, personality traits, beliefs, preference, relationships, attributes, goals, and experiences).} \\
        \texttt{\hspace{1em}- A response by the listener which makes a specific choice among the given options in the format specified after \#\#\# Format.} \\[1em]

    \texttt{Please solve the above task step by step as follows:} \\
    \texttt{Step 1. Find the next question utterance in the given dialogue, and continue to Step 2 if found; otherwise, exit.} \\
    \texttt{Step 2. Judge whether the question found in Step 1 asks a single listener to choose among specific options, continue to Step 3 if yes; otherwise, go to Step 1.} \\
        \texttt{\hspace{1em}- Questions without specific options being presented should be excluded (\textit{e.g.}, "where are you going?")} \\
    \texttt{Step 3. Judge whether the choice would be influenced by the listener's personas, continue to Step 4 if yes; otherwise, go to Step 1.} \\
        \texttt{\hspace{1em}- The question that directly asks a piece of persona should be excluded (\textit{e.g.}, "Do you like apples?")} \\
        \texttt{\hspace{1em}- The question that directly asks a fact should be excluded (\textit{e.g.}, “Did you get the cookie?”)} \\
    \texttt{Step 4. Find the nearest response by the listener in which the listener makes a specific choice among the given options, return the two utterances as pairs if found; otherwise, go to Step 1.} \\[1em]
    
    \texttt{\#\#\# Format:} \\
    \texttt{Each utterance in the dialogue has a unique numeric identifier (\textit{e.g.}, 1, 2, 3...).} \\
    \texttt{Return your result as a list of **(question, response)** number pairs:} \\
    \texttt{[(3, 5), (8, 9), ...]} \\
    
    \bottomrule
  \end{tabular}
  \caption{Prompt for detecting potential PICQs and their corresponding answers.}
  \label{tab:step1_gpt}
\end{table*}

\begin{table*}[t] 
  \footnotesize
  \begin{tabular}{p{0.97\linewidth}} 
    \toprule
    \texttt{Please review the specific question-answering pairs identified by GPT-4.1 and the whole scene dialogue, then judge whether each pair is correct or incorrect based on the guidelines below.} \\[1em]
    
    \texttt{Task:} \\
    \texttt{For each pair (question utterance, answer) determine whether:} \\
    \texttt{1. Whether the question asks a single listener to choose among specific options} \\
        \texttt{\hspace{1em} a. Question without specific options is incorrect (\textit{e.g.}, "where are you going?")} \\
    \texttt{2. Whether the choice would be influenced by the listener's personas (\textit{e.g.}, personality traits, beliefs, preference, relationships, attributes, goals, and experiences)} \\
        \texttt{\hspace{1em}a. Question that directly asks a piece of persona incorrect (\textit{e.g.}, "Do you like apples?")} \\
        \texttt{\hspace{1em}b. Question that directly asks a fact is incorrect (\textit{e.g.}, “Did you get the cookie?”)} \\
    \texttt{3. Whether the answer is the nearest response by the listener in which the listener makes a specific choice among the given options} \\[1em]
    
    \texttt{If the pair satisfies all the above requirements, label correct, otherwise incorrect.} \\
    \bottomrule
  \end{tabular}
  \caption{Instruction for human annotators to check potential PICQs and their corresponding answers.}
  \label{tab:step1_human}
\end{table*}

\begin{table*}[t] 
  \footnotesize
  \begin{tabular}{p{0.97\linewidth}} 
    \toprule
    \texttt{Given dialogue context, question utterance, and basic persona (age, gender, and basic relationship with the questioner) of the listener, your task is to:} \\[1em]
    \texttt{Conduct query-focused summarization based on the question and dialogue context. The summary should be self-contained, allowing someone to understand the listener's choice-making situation (what they need to choose and why) just by reading the summary, without referring back to the original dialogue.} \\
        \texttt{\hspace{1em}a. If the dialogue context and question do not contain enough specific information about the listener's choice-making situation (or are not self-contained), skip them.} \\   
    \bottomrule
  \end{tabular}
  \caption{Instruction for query-focused summarization.}
  \label{tab:step2_summarization}
\end{table*}

\begin{table*}[t] 
  \scriptsize
  \begin{tabular}{p{0.97\linewidth}} 
    \toprule
    \texttt{Given dialogue context, question utterance, and basic persona (age, gender, and basic relationship with the questioner) of the listener, your task is to:} \\[1em]
    
    \texttt{Identify missing persona (not explicitly stated anywhere in the provided context and basic persona) of the listener that will influence the choice of the options provided in the question the most.} \\
        \texttt{\hspace{1em} a. You should identify up to five pieces of missing persona. Prioritize based on factors that:} \\
            \texttt{\hspace{2em}i. Represent strong motivations or driving forces behind the choice (\textit{e.g.}, key personal goals, deeply held beliefs, very strong preferences).} \\
            \texttt{\hspace{2em}ii. Act as necessary conditions, core constraints, or essential enablers (\textit{e.g.}, affordability for a large purchase, prerequisite required skills).} \\
            \texttt{\hspace{2em}iii. Are critical factors when their value falls within a certain range (\textit{e.g.}, “spice tolerance” when choosing a Sichuan (spicy) vs. Japanese restaurant).} \\
        \texttt{\hspace{1em} b. When you consider each piece of the required persona, you should first choose a category and choose the specific linguistic patterns associated with the category to describe the specific persona required to make a choice. You can write more than one piece of the persona for the same category, and you can skip some categories if they are irrelevant. We count the number of pieces of the required persona by the number of descriptions you have provided (different pieces of persona in the same category should be treated):} \\
            \texttt{\hspace{2em}i. Personality: } \\
                \texttt{\hspace{3em}1. whether s/he is ADJ (ADJ is an adjective describing a personality trait); for example,} \\
                    \texttt{\hspace{4em}- whether s/he is introverted} \\
                    \texttt{\hspace{4em}- whether s/he is adventurous} \\
            \texttt{\hspace{2em}ii. Beliefs (personal values, moral principles, and views on social norms): } \\
                \texttt{\hspace{3em}1. whether s/he believes it's ADJ to VP (ADJ is an adjective to comment a behavior, VP is a verb phrase); for example,} \\
                    \texttt{\hspace{4em}- whether s/he believes it's important to save money} \\
                    \texttt{\hspace{4em}- whether s/he believes it's wrong to lie to others} \\
                \texttt{\hspace{3em}2. whether s/he believes propN should VP (propN is a target, VP is a verb phrase); for example, } \\
                    \texttt{\hspace{4em}- whether s/he believes children should have less screen time} \\
                    \texttt{\hspace{4em}- whether s/he believes the government should invest more in public transport} \\
            \texttt{\hspace{2em}iii. Tastes:} \\
                \texttt{\hspace{3em}1. whether s/he (dis)likes VP (VP is a verb phrase); for example,} \\
                    \texttt{\hspace{4em}- whether s/he likes traveling  } \\
                    \texttt{\hspace{4em}- whether s/he dislikes waking up early } \\
                \texttt{\hspace{3em}2. whether s/he (dis)likes N (N is a noun); for example,} \\
                    \texttt{\hspace{4em}- whether s/he likes spicy food} \\
                    \texttt{\hspace{4em}- whether s/he dislikes crowded places } \\
            \texttt{\hspace{2em}iv. Relationships:} \\
                \texttt{\hspace{3em}1. whether s/he is N of propN (N is a noun representing a human relationship, propN is a name of a person); for example,} \\
                    \texttt{\hspace{4em}- whether s/he is a close friend of Alex } \\
                    \texttt{\hspace{4em}- whether s/he is the sibling of Sarah } \\
                \texttt{\hspace{3em}2. whether s/he is ADJ + P + propN (ADJ represents an adjective or a past participle used adjectivally, describes the subject's (s/he's) view, attitude, feeling, or judgment regarding the person propN, P is prepositional); for example,} \\
                    \texttt{\hspace{4em}- whether s/he is annoyed with Maria} \\
                    \texttt{\hspace{4em}- whether s/he is loyal to their team} \\
            \texttt{\hspace{2em}v. Attributes:} \\
                \texttt{\hspace{3em}1. whether his/her ATTR is X  (ATTR is a noun describing an attribute of a person, (\textit{e.g.}, gender, occupation, age, height, weight, income, etc.) X is a specific attribute value or an expression describing a range of values); for example,} \\
                    \texttt{\hspace{4em}- whether his/her physical stamina is suitable for a long hike} \\
                    \texttt{\hspace{4em}- whether his/her disposable income is suitable for luxury purchases} \\
            \texttt{\hspace{2em}vi. Goals (short-term or long-term goals):} \\
                \texttt{\hspace{3em}1. whether s/he aims to VP (VP is a verb phrase describing the goal); for example,} \\
                    \texttt{\hspace{4em}- whether s/he aims to get a promotion } \\
                    \texttt{\hspace{4em}- whether s/he aims to learn a new language } \\
            \texttt{\hspace{2em}vii. Experience (Write 'experience' only if the specific past event memory itself is directly influencing the choice, rather than having been internalized as a taste or other categories):  } \\
                \texttt{\hspace{3em}1. whether s/he has V (V is a past participle phrase); for example,} \\
                    \texttt{\hspace{4em}- whether s/he has been to that restaurant before} \\
                    \texttt{\hspace{4em}- whether s/he has had a bad experience with online shopping } \\
        \texttt{\hspace{1em} c. After identifying a specific piece of persona that significantly influences the choice, consider if it can be expressed in a more generalized or abstract way without losing its core impact or clarity. Meanwhile, prioritize annotating plausible and impactful missing personas, avoiding overly specific or highly improbable scenarios unless contextually supported; for example,} \\
            \texttt{\hspace{2em}- whether s/he likes Indian chicken curry -> whether s/he likes curry} \\
            \texttt{\hspace{2em}- whether s/he is helpful in park at night -> whether s/he is helpful} \\
        \texttt{\hspace{1em} d. Apart from using "s/he" to refer to the respondent, do not use pronominal reference for other entities, even if they are present in the context.} \\
  
    \bottomrule
  \end{tabular}
  \caption{Instruction for identifying missing relevant personas.}
  \label{tab:step2_identify}
\end{table*}

\begin{table*}[t] 
  \scriptsize
  \begin{tabular}{p{0.97\linewidth}} 
    \toprule
    \texttt{Given dialogue context, question utterance, and basic persona (age, gender, and basic relationship with the questioner) of the listener, your task is to:} \\[1em]
    
    \texttt{Identify missing persona (not explicitly stated anywhere in the provided context and basic persona) of the listener that will influence the choice of the options provided in the question the most.} \\
        \texttt{\hspace{1em} a. You should identify up to five pieces of missing persona. Prioritize based on factors that:} \\
            \texttt{\hspace{2em}i. Represent strong motivations or driving forces behind the choice (\textit{e.g.}, key personal goals, deeply held beliefs, very strong preferences).} \\
            \texttt{\hspace{2em}ii. Act as necessary conditions, core constraints, or essential enablers (\textit{e.g.}, affordability for a large purchase, prerequisite required skills).} \\
            \texttt{\hspace{2em}iii. Are critical factors when their value falls within a certain range (\textit{e.g.}, “spice tolerance” when choosing a Sichuan (spicy) vs. Japanese restaurant).} \\
        \texttt{\hspace{1em} b. When you consider each piece of the required persona, you should first choose a category and choose the specific linguistic patterns associated with the category to describe the specific persona required to make a choice. You can write more than one piece of the persona for the same category, and you can skip some categories if they are irrelevant. We count the number of pieces of the required persona by the number of descriptions you have provided (different pieces of persona in the same category should be treated):} \\
            \texttt{\hspace{2em}i. Personality: } \\
                \texttt{\hspace{3em}1. whether s/he is ADJ (ADJ is an adjective describing a personality trait); for example,} \\
                    \texttt{\hspace{4em}- whether s/he is introverted} \\
                    \texttt{\hspace{4em}- whether s/he is adventurous} \\
            \texttt{\hspace{2em}ii. Beliefs (personal values, moral principles, and views on social norms): } \\
                \texttt{\hspace{3em}1. whether s/he believes it's ADJ to VP (ADJ is an adjective to comment a behavior, VP is a verb phrase); for example,} \\
                    \texttt{\hspace{4em}- whether s/he believes it's important to save money} \\
                    \texttt{\hspace{4em}- whether s/he believes it's wrong to lie to others} \\
                \texttt{\hspace{3em}2. whether s/he believes propN should VP (propN is a target, VP is a verb phrase); for example, } \\
                    \texttt{\hspace{4em}- whether s/he believes children should have less screen time} \\
                    \texttt{\hspace{4em}- whether s/he believes the government should invest more in public transport} \\
            \texttt{\hspace{2em}iii. Tastes:} \\
                \texttt{\hspace{3em}1. whether s/he (dis)likes VP (VP is a verb phrase); for example,} \\
                    \texttt{\hspace{4em}- whether s/he likes traveling  } \\
                    \texttt{\hspace{4em}- whether s/he dislikes waking up early } \\
                \texttt{\hspace{3em}2. whether s/he (dis)likes N (N is a noun); for example,} \\
                    \texttt{\hspace{4em}- whether s/he likes spicy food} \\
                    \texttt{\hspace{4em}- whether s/he dislikes crowded places } \\
            \texttt{\hspace{2em}iv. Relationships:} \\
                \texttt{\hspace{3em}1. whether s/he is N of propN (N is a noun representing a human relationship, propN is a name of a person); for example,} \\
                    \texttt{\hspace{4em}- whether s/he is a close friend of Alex } \\
                    \texttt{\hspace{4em}- whether s/he is the sibling of Sarah } \\
                \texttt{\hspace{3em}2. whether s/he is ADJ + P + propN (ADJ represents an adjective or a past participle used adjectivally, describes the subject's (s/he's) view, attitude, feeling, or judgment regarding the person propN, P is prepositional); for example,} \\
                    \texttt{\hspace{4em}- whether s/he is annoyed with Maria} \\
                    \texttt{\hspace{4em}- whether s/he is loyal to their team} \\
            \texttt{\hspace{2em}v. Attributes:} \\
                \texttt{\hspace{3em}1. whether his/her ATTR is X  (ATTR is a noun describing an attribute of a person, (\textit{e.g.}, gender, occupation, age, height, weight, income, etc.) X is a specific attribute value or an expression describing a range of values); for example,} \\
                    \texttt{\hspace{4em}- whether his/her physical stamina is suitable for a long hike} \\
                    \texttt{\hspace{4em}- whether his/her disposable income is suitable for luxury purchases} \\
            \texttt{\hspace{2em}vi. Goals (short-term or long-term goals):} \\
                \texttt{\hspace{3em}1. whether s/he aims to VP (VP is a verb phrase describing the goal); for example,} \\
                    \texttt{\hspace{4em}- whether s/he aims to get a promotion } \\
                    \texttt{\hspace{4em}- whether s/he aims to learn a new language } \\
            \texttt{\hspace{2em}vii. Experience (Write 'experience' only if the specific past event memory itself is directly influencing the choice, rather than having been internalized as a taste or other categories):  } \\
                \texttt{\hspace{3em}1. whether s/he has V (V is a past participle phrase); for example,} \\
                    \texttt{\hspace{4em}- whether s/he has been to that restaurant before} \\
                    \texttt{\hspace{4em}- whether s/he has had a bad experience with online shopping } \\

            \texttt{Output strictly in the following format:} \\
            \texttt{(personality) whether he is introverted} \\
        \texttt{(tastes) whether she dislikes waking up early.} \\[1em]
    
        \texttt{Do not output additional explanation!} \\
        \texttt{Here is the basic persona about \{choice-maker\}: \{basic\_info\}} \\
        \texttt{The following is the conversation, with the final utterance being the question utterance:} \\
    \bottomrule
  \end{tabular}
  \caption{Prompt for identifying missing relevant personas.}
  \label{tab:prompt_identify}
\end{table*}

\begin{table*}[t] 
  \scriptsize
  \begin{tabular}{p{0.97\linewidth}} 
    \toprule
    \texttt{Given dialogue context, question utterance, and basic persona (age, gender, and basic relationship with the questioner) of the listener, your task is to:} \\[1em]
    
    \texttt{Identify missing persona (not explicitly stated anywhere in the provided context and basic persona) of the listener that will influence the choice of the options provided in the question most.} \\
        \texttt{\hspace{1em} a. You should identify up to five pieces of missing persona. Prioritize based on factors that:} \\
            \texttt{\hspace{2em}i. Represent strong motivations or driving forces behind the choice (\textit{e.g.}, key personal goals, deeply held beliefs, very strong preferences).} \\
            \texttt{\hspace{2em}ii. Act as necessary conditions, core constraints, or essential enablers (\textit{e.g.}, affordability for a large purchase, prerequisite required skills).} \\
            \texttt{\hspace{2em}iii. Are critical factors when their value falls within a certain range (\textit{e.g.}, “spice tolerance” when choosing a Sichuan (spicy) vs. Japanese restaurant).} \\
        \texttt{\hspace{1em} b. When you consider each piece of the required persona, you should first choose a category and choose the specific linguistic patterns associated with the category to describe the specific persona required to make a choice. You can write more than one piece of the persona for the same category, and you can skip some categories if they are irrelevant. We count the number of pieces of the required persona by the number of descriptions you have provided (different pieces of persona in the same category should be treated):} \\
            \texttt{\hspace{2em}i. Personality: } \\
                \texttt{\hspace{3em}1. whether s/he is ADJ (ADJ is an adjective describing a personality trait); for example,} \\
                    \texttt{\hspace{4em}- whether s/he is introverted} \\
                    \texttt{\hspace{4em}- whether s/he is adventurous} \\
            \texttt{\hspace{2em}ii. Beliefs (personal values, moral principles, and views on social norms): } \\
                \texttt{\hspace{3em}1. whether s/he believes it's ADJ to VP (ADJ is an adjective to comment a behavior, VP is a verb phrase); for example,} \\
                    \texttt{\hspace{4em}- whether s/he believes it's important to save money} \\
                    \texttt{\hspace{4em}- whether s/he believes it's wrong to lie to others} \\
                \texttt{\hspace{3em}2. whether s/he believes propN should VP (propN is a target, VP is a verb phrase); for example, } \\
                    \texttt{\hspace{4em}- whether s/he believes children should have less screen time} \\
                    \texttt{\hspace{4em}- whether s/he believes the government should invest more in public transport} \\
            \texttt{\hspace{2em}iii. Tastes:} \\
                \texttt{\hspace{3em}1. whether s/he (dis)likes VP (VP is a verb phrase); for example,} \\
                    \texttt{\hspace{4em}- whether s/he likes traveling  } \\
                    \texttt{\hspace{4em}- whether s/he dislikes waking up early } \\
                \texttt{\hspace{3em}2. whether s/he (dis)likes N (N is a noun); for example,} \\
                    \texttt{\hspace{4em}- whether s/he likes spicy food} \\
                    \texttt{\hspace{4em}- whether s/he dislikes crowded places } \\
            \texttt{\hspace{2em}iv. Relationships:} \\
                \texttt{\hspace{3em}1. whether s/he is N of propN (N is a noun representing a human relationship, propN is a name of a person); for example,} \\
                    \texttt{\hspace{4em}- whether s/he is a close friend of Alex } \\
                    \texttt{\hspace{4em}- whether s/he is the sibling of Sarah } \\
                \texttt{\hspace{3em}2. whether s/he is ADJ + P + propN (ADJ represents an adjective or a past participle used adjectivally, describes the subject's (s/he's) view, attitude, feeling, or judgment regarding the person propN, P is prepositional); for example,} \\
                    \texttt{\hspace{4em}- whether s/he is annoyed with Maria} \\
                    \texttt{\hspace{4em}- whether s/he is loyal to their team} \\
            \texttt{\hspace{2em}v. Attributes:} \\
                \texttt{\hspace{3em}1. whether his/her ATTR is X  (ATTR is a noun describing an attribute of a person, (\textit{e.g.}, gender, occupation, age, height, weight, income, etc.) X is a specific attribute value or an expression describing a range of values); for example,} \\
                    \texttt{\hspace{4em}- whether his/her physical stamina is suitable for a long hike} \\
                    \texttt{\hspace{4em}- whether his/her disposable income is suitable for luxury purchases} \\
            \texttt{\hspace{2em}vi. Goals (short-term or long-term goals):} \\
                \texttt{\hspace{3em}1. whether s/he aims to VP (VP is a verb phrase describing the goal); for example,} \\
                    \texttt{\hspace{4em}- whether s/he aims to get a promotion } \\
                    \texttt{\hspace{4em}- whether s/he aims to learn a new language } \\
            \texttt{\hspace{2em}vii. Experience (Write 'experience' only if the specific past event memory itself is directly influencing the choice, rather than having been internalized as a taste or other categories):  } \\
                \texttt{\hspace{3em}1. whether s/he has V (V is a past participle phrase); for example,} \\
                    \texttt{\hspace{4em}- whether s/he has been to that restaurant before} \\
                    \texttt{\hspace{4em}- whether s/he has had a bad experience with online shopping } \\

        \texttt{\hspace{2em}3. After identifying a specific piece of persona that significantly influences the decision, consider if it can be expressed in a more generalized or abstract way without losing its core impact or clarity to avoid overly specific or highly improbable scenarios unless contextually supported. For example,  } \\                  
                    \texttt{\hspace{3em}- whether s/he likes Indian chicken curry -> whether s/he likes curry} \\
                    \texttt{\hspace{3em}- whether s/he is helpful in park at night -> whether s/he is helpful} \\
        
        \texttt{\hspace{2em}4. You should first summarize what choice {maker} is requested to make and then identify the missing persona, finally generalize them} \\
    
    \texttt{Output strictly in the following format:} \\
    \texttt{(summary) ...} \\
    \texttt{(personality) whether he is introverted} \\
    \texttt{(tastes) whether he likes Indian chicken curry} \\
    \texttt{...} \\
    \texttt{[generalized]} \\
    \texttt{(personality) whether he is introverted} \\
    \texttt{(tastes) whether he likes curry} \\
    \texttt{...} \\
    
    \texttt{Do not output additional explanation!} \\
    \texttt{Here is the basic persona about \{choice-maker\}: \{basic\_info\}} \\
    \texttt{The following is the conversation, with the final utterance being the question utterance:} \\
    \bottomrule
  \end{tabular}
  \caption{Prompt for identifying missing personas in multi-task instruction strategy.}
  \label{tab:prompt_identifycot}
\end{table*}

\begin{table*}[t] 
  \footnotesize
  \begin{tabular}{p{0.97\linewidth}} 
    \toprule
        \texttt{Given a summary showing the target's choice-making situation (what they need to choose and why) and their basic persona and a list of persona dimensions (meaning the specific value of that persona for the target is currently unknown), your task is to assign an influence score for each piece of persona dimension considering their possible influence on the choice-making.} \\[1em]
        
        \texttt{Definition of influence score:} \\[1em]
        
        \texttt{Influence refers to how strongly a piece of persona dimension affects the answer to the question. It is rated as follows:} \\[1em]
        \texttt{Score 0 - Irrelevant} \\
        \texttt{The persona dimension, regardless of its value or state, has no influence on the choice outcome.} \\
        \texttt{\textit{e.g.}, “favorite color” generally has no impact on deciding whether to accept a job offer.} \\[1em]
        \texttt{Score 1 - Minor Influence} \\
        \texttt{The persona dimension has some influence on the choice-making process or outcome, but this influence is not strong enough to be considered key or decisive. It might affect preferences for details, execution, or make one option slightly more or less appealing, but it does not fundamentally drive or constrain the core choice.} \\
        \texttt{\textit{e.g.}, "A slight preference for Software X's user interface over Software Y's, when both tools meet all core functional requirements and are within budget," might influence which tool the team adopts, but it would not lead them to choose Software X if it lacked a critical feature that Software Y possessed.} \\[1em]
        \texttt{Score 2 - Key Influence} \\
        \texttt{The persona dimension is a key factor that influences, changes, or determines the main choice outcome. This influence can manifest as a necessary condition/constraint/enabler (making an option impossible or essential under certain conditions) OR as a strong motivation that shapes the choice.} \\
        \texttt{\textit{e.g.}, "Spice tolerance" is key if extremely low (effectively vetoes Sichuan, acting as a constraint). "Having a driver's license" is key if the job requires driving (a necessary condition).} \\
        \texttt{\textit{e.g.}, "whether s/he likes alcohol" significantly influences a choice about drinking beer (as a strong motivation) whether s/he likes or dislikes alcohol.} \\[1em]
        
        \texttt{You should first consider the possible values of each piece of missing personal information and then judge its score.} \\[1em]
        
        \texttt{Summary, basic persona, and missing persona dimensions are given as follows:} \\

    \bottomrule
  \end{tabular}
  \caption{Prompt for calculating the influence score for missing relevant personas.}
  \label{tab:prompt_Influence}
\end{table*}

\begin{table*}[t] 
  \footnotesize
  \begin{tabular}{p{0.97\linewidth}} 
    \toprule

        \texttt{Your task is to analyze the persona dimensions (meaning the specific value of that persona for the target is currently unknown) from the perspective of the target individual based on the provided Basic Persona.}\\
        \texttt{You will receive the persona dimensions belonging to that same individual.}\\
        \texttt{You should assign an inaccessibility score (0-2) to each Info. This score reflects the individual's willingness to disclose that specific persona dimension, based on the benchmark of who is proactively asking them about it.}\\[1em]
        \texttt{Scoring Scale:}\\

        \texttt{Score 0: Public / Observable Information}\\
        \texttt{Willingness Benchmark: The individual would share this with a stranger, or the information is physically observable/public knowledge anyway.}\\
        \texttt{Examples: name, visible appearance (hair color, height), accent.}\\[1em]
        \texttt{Score 1: General Acquaintance Information}\\
        \texttt{Willingness Benchmark: The individual would NOT share this with a random stranger, but would comfortably share it with a general acquaintance (like a co-worker, a neighbor, or a casual friend) if the topic came up in conversation.}\\
        \texttt{Examples: Job title, general hobbies, hometown, favorite sports team, where s/he went to college.}\\[1em]
        \texttt{Score 2: Close Acquaintance Information}\\
        \texttt{Willingness Benchmark: The individual would only share this information with a close acquaintance whom they trust (such as a close friend, immediate family, or a spouse). They would actively avoid discussing this with co-workers or casual friends. }\\
        \texttt{Examples: specific salary/financial struggles, private family issues, detailed medical conditions, strong political opinions, deep life aspirations.}\\[1em]

        \texttt{You may first imagine you are the individual and then consider whether it is comfortable for you to share the personal information during the conversation with a certain group of people.}\\

        \texttt{Basic persona description and persona dimensions are given as follows:}\\
       
    \bottomrule
  \end{tabular}
  \caption{Prompt for calculating the inaccessibility score for missing relevant personas.}
  \label{tab:prompt_Difficulty}
\end{table*}

\end{document}